\documentclass[5p]{elsarticle}
 \usepackage{graphicx, color}
\usepackage{lineno,hyperref}
\usepackage{here}
\modulolinenumbers[2]

\begin{document}

\begin{frontmatter}

\title{Development of an ``$\alpha$-ToF'' detector for correlated measurement of atomic masses
        \\ and decay properties}

\author[add2,add1,add3]{T.~Niwase\corref{mycorrespondingauthor}}
\cortext[mycorrespondingauthor]{Corresponding author}

\ead{toshitaka.niwase@riken.jp}

\author[add3]{M.~Wada}
\author[add3]{P.~Schury}
\author[add1]{H.~Haba}
\author[add1,add5,add3]{S.~Ishizawa}
\author[add4]{Y.~Ito}
\author[add1]{D.~Kaji}
\author[add1]{S.~Kimura}
\author[add3]{H.~Miyatake}
\author[add1]{K.~Morimoto}
\author[add2,add1]{K.~Morita}
\author[add1]{M.~Rosenbusch}
\author[add6]{H.~Wollnik}
\author[add7]{T.~Shanley}
\author[add7]{Y.~Benari}

\address[add2]{Department of Physics, Kyushu University, Fukuoka City, Fukuoka, Japan}
\address[add1]{RIKEN, Nishina Center for Accelerator-Based Science, Wako City, Saitama, Japan}
\address[add3]{Wako Nuclear Science Center(WNSC), Institute of Particle and Nuclear Studies(IPNS),\\
High Energy Accelerator Research Organization(KEK), Wako City, Saitama, Japan}
\address[add4]{Japan Atomic Energy Agency(JAEA), Tokai, Ibaraki, Japan}
\address[add5]{Graduate School of Science and Engineering, Yamagata University, Yamagata, Japan}
\address[add6]{Department of Chemistry and Biochemistry, New Mexico State University, Las Cruces, NM, USA}
\address[add7]{ETP Ion Detect, 8 Martha Street, Clyde, New South Wales, Australia}

\begin{abstract}
  \par
We have developed a novel detector, referred to as an ``$\alpha$-ToF detector'', for correlated measurements of atomic masses and decay properties of low-yield, short-lived radioactive isotopes using a multi-reflection time-of-flight mass spectrograph.  By correlating measured time-of-flight signals with decay events, it will be possible to suppress background events and obtain accurate, high-precision mass values even in cases of very low event rates. An offline test of the $\alpha$-ToF detector has shown that the time-of-flight detection efficiency for 5.48~MeV $\alpha$-rays is more than 90\% and yields a time resolution of 251.5(68)~ps and an energy resolution of 141.1(9)~keV.  Using a two-dimensional spectrum of the correlated $\alpha$-ray energy and time-of-flight, the $\alpha$-rays from mixed $\alpha$ sources could be fairly well resolved.

\end{abstract}


\end{frontmatter}


\section{Introduction}
The multi-reflection time-of-flight mass spectrograph (MRTOF-MS) was invented nearly 30 years ago for general purpose mass measurements \cite{HWollnik1990}. In 2013, such devices were used to measure the masses of radioactive nuclei at three different laboratories for the first time \cite{YIto2013011306,WolfMTOF2013123,PlassMTOF2013134}. Since then, the device has been acknowledged to be useful 
for mass measurement of short-lived rare isotopes. We performed mass measurements of fusion-evaporation products \cite{PSchury2017011305,SKimura2018134,MRosenbusch2018064306,YIto2018152501} provided at the RIKEN gas-filled recoil ion separator GARIS-II \cite{DKaji2013311} and showed that an MRTOF-MS is applicable to short-lived nuclei having a half-life of $\geq$ 10 ms by achieving a relative mass precision of $\delta$$m$/$m$$\sim$10$^{-7}$ \cite{MWadaAPR2017}, and to very low-yield isotopes by mass analyzing $^{249}$Md which had a detection rate of one ion per 1000 seconds \cite{YIto2018152501}.

\par We plan to measure even lower-yield isotopes, such as superheavy nuclei which will have 
yield of one ion per day or less. In such a measurement, reliable discrimination of background events is essential.  Decay properties -- such as emitted particles and their energies, along with decay half-lives -- can be used as fingerprints of  radioactive isotopes. If we can correlate such decay properties with the time-of-flight (ToF) signal, we will be able to distinguish a true event from background events which can originate from stable, possibly molecular, contaminant ions as well as dark counts in the detector.
 
\par For this purpose, we have developed a new detector, referred to as an ``$\alpha$-ToF detector'', which can measure an ion's time-of-flight as well as its decay energy and absolute time of subsequent decays from implanted radioactive ions.  In addition to background reduction in mass measurements, this detector will be used for nuclear spectroscopy.

\section{Apparatus}

\begin{figure}[h]
\begin{center}
\includegraphics[scale=0.45]{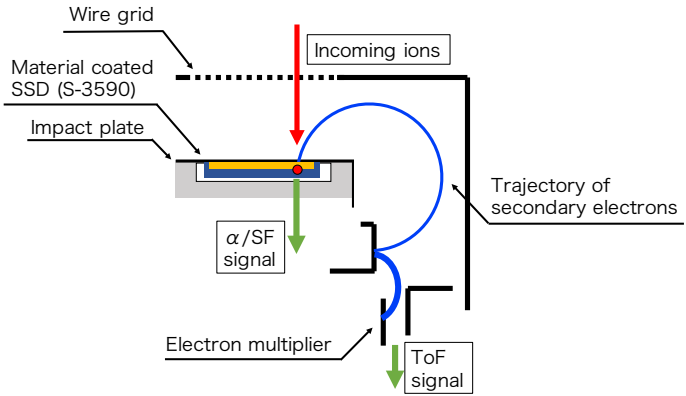}
\end{center}
\caption{Schematic view of the $\alpha$-ToF detector.}
\label{fig1}
\end{figure}

The $\alpha$-ToF detector is a modified commercial MagneToF detector (ETP 14DM572), as shown in Fig.\ \ref{fig1}. It consists of an impact plate, an internal circuit for generating a suitable potentials, permanent magnets, and an electron multiplier. When an ion hits the impact plate of the MagneToF detector, secondary electrons are emitted and those electrons are isochronously guided by crossed electric and magnetic fields to the electron multiplier amplifying section of the detector to provide a fast timing signal for the ions.
We replaced the impact plate with a silicon surface barrier detector (SSD, Hamamatsu S-3590) whose surface was coated for efficient secondary electron emission, where both Au+Al$_2$O$_3$ and Au+MgO were tested.

\par
\begin{figure}[H]
\begin{center}
\includegraphics[scale=0.44]{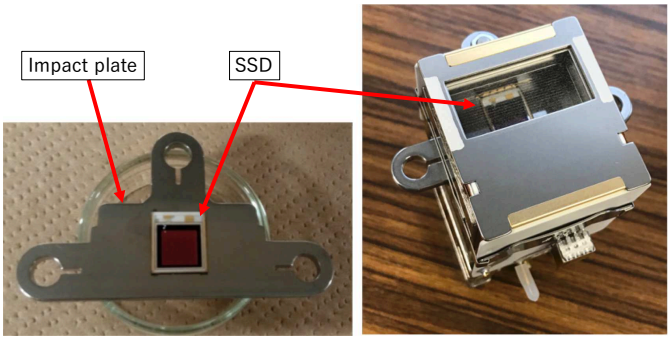}
\end{center}
\caption{Photographs showing the impact plate with SSD (left) and the assembled $\alpha$-ToF detector (right)}
\label{fig2}
\end{figure}

 In order to maintain the potential of the impact plate, the gold coated surface area was wire-bonded to the terminal of the front side of the SSD, which is electrically connected to the impact plate. MgO and Al$_{2}$O$_{3}$ are materials known to have high secondary electron emission probabilities \cite{JohnsonSE1953582,UshioSE1988299,NilssonSE1970301} and no deliquescence. The gold and Al$_{2}$O$_{3}$ coatings were made by a vacuum evaporation method, while the MgO coating was made by a sputtering method.  The thickness of each layer is roughly ten nanometers. Photographs of the impact plate and the $\alpha$-ToF detector assembly are shown in Fig.\ref{fig2}.

\par
\begin{figure}[t]
\begin{center}
\includegraphics[scale=0.45]{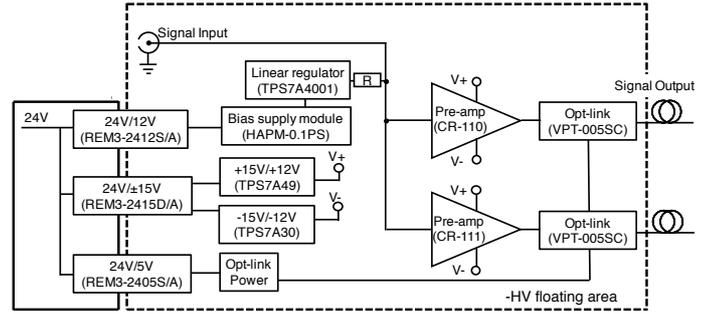}
\end{center}
\caption{Sketch of the floating signal processing front-end circuitry. The area surrounded by the broken line is on the high voltage ground plane. This circuit consists of a high-gain preamplifier (Cremat CR-110, 1.6~V/pC) and a low-gain preamplifier (Cremat CR-111, 0.16~V/pC), 
 ultra low noise voltage regulators (Texas Instruments, TPS7A49/30) for the preamplifiers, a -70 V power supply for the SSD bias (Matsusada, HAPM-0.1PS) with high voltage linear regulator (Texas Instruments, TPS7A4001), and optically isolated ``Opt-link" signal transceivers (Nanaboshi, VPT(R)-005SC). 
The electric power for the isolated circuits are supplied by DC/DC converters with 5~kV isolation (RECOM, REM3 series).
 }
\label{fig3}
\end{figure}

 In normal operation, the $\alpha$-ToF detector's impact plate must be placed at a negative high potential, approximately $-2$ kV. Since the SSD is also placed on the impact plate potential, an optically isolated circuit was developed to provide a bias voltage to the SSD, and to amplify the decay signal.
A sketch of the design for such an optical isolation system is shown in Fig.\ \ref{fig3}. The signal from the Si detector is amplified by two charge sensitive pre-amplifiers having different gains: a high-gain one for $\alpha$-ray signals and a low-gain one for spontaneous fission signals. The amplified analog signals are sent to shaping amplifiers on ground potential through optical transceivers. The data from the Si detector and the time-of-flight data are separately recorded event by event with absolute time stamps.

\section{Offline performance tests}
\par 
The new detector was thoroughly tested offline. In particular, the detection efficiency and timing resolution were confirmed to be sufficient to perform atomic mass measurements. In addition, the new detector's fitness for particle identification was confirmed by $\alpha$-energy correlated time-of-flight ($E_{\alpha}$-ToF) measurements.
 
\subsection{Detection efficiency}
\begin{figure}[H]
\begin{center}
\includegraphics[scale=0.65]{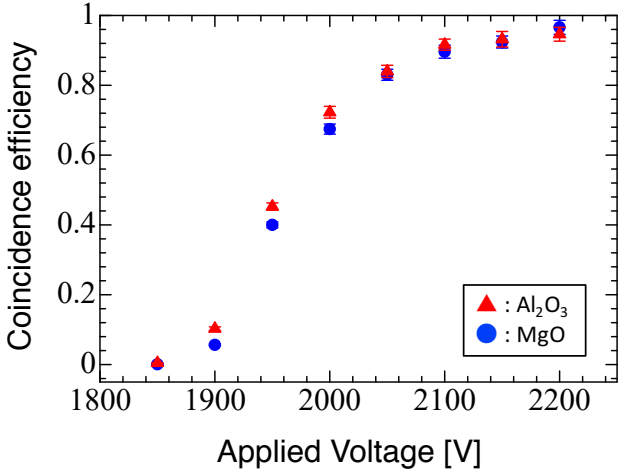}
\end{center}
\caption{Detection efficiencies measured for 5.48 MeV $\alpha$-rays as functions of impact plate potential. Red triangles and blue circles designate results from Al$_{2}$O$_{3}$ and MgO coatings, respectively. \label{fig4}
}

\end{figure}

We measured the detection efficiency of the $\alpha$-ToF detector by using collimated 5.48 MeV $\alpha$-rays from a $^{241}$Am source. As the SSD should have near unity detection efficiency for 5.48 MeV $\alpha$-rays, the efficiency was defined as the ratio of coincident ToF signals to the $\alpha$-singles counts. The counting rate was set to be as low as $\approx$0.1 cps by using a collimator.  A ToF signal was considered to be coincident when the timing signal were detected within 500~ns of each other.  The $\alpha$-singles signals were limited to those signals with an SSD signal amplitude commensurate with $^{241}$Am $\alpha$-rays.  Fig.~\ref{fig4} shows the detection efficiency measured for both MgO and Al$_{2}$O$_3$ surface coatings as functions of the high-voltage potential applied to the impact plate.  At the nominal operation potential of $-2100$~V, the efficiency was more than 90\% for both surface coatings.

\subsection{Time resolution}
\begin{figure}[H]
 \begin{center}
  \includegraphics[scale=0.45]{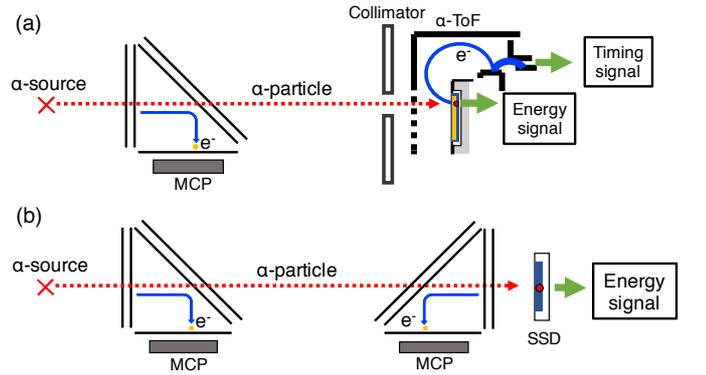}
  \caption{(a) Setup for configuration used in the offline tests. For detection efficiency and time resolution measurements, $^{241}$Am source was used, while for correlation measurement, a mixed $\alpha$ source was used. (b) Comparison setup to determine the intrinsic time resolution of a triangle-roof detector (see text.) \label{fig5}}

 \end{center}
\end{figure}

The reference timing signal for the time-of-flight was provided by a Busch type triangle-roof time-of-flight detector \cite{FBusch198071,KMorimotoAPR2012} located 27 cm upstream from the $\alpha$-ToF detector as shown in Fig.\ \ref{fig5}(a)
The time resolution was determined from the time-of-flight spectrum.
In the following discussion, the time resolution of  the triangle-roof detector and the $\alpha$-ToF detector are indicated as $\sigma_{\rm{tri}}$ and $\sigma_{\alpha{\rm ToF}}$ , respectively. The measured time resolution, being a convolution of these two, is written $\sigma_{{\rm tri}-\alpha{\rm ToF}}$.

As the start signal was given by the triangle-roof detector and the stop signal by the $\alpha$-ToF detector, the measured time resolution, $\sigma_{{\rm tri}-\alpha{\rm ToF}}=413.7(33)$~ps. However, as this includes both the start timing of the triangle-roof detector and the stop timing in the $\alpha$-ToF, a comparison setup has been used as shown in Fig.\ \ref{fig5}(b). Two triangle-roof detectors of the same type have been placed opposite to each other, where one has been used as the start and the other as the stop signal. An SSD was placed behind the triangle-roof detector to produce a trigger signal as in the detection efficiency measurement. The standard deviation of the time differences was found to be
$\sigma_{{\rm tri}-{\rm tri}} =$ 465.5(40)~ps.
In that way, the resolution of a single triangle-roof detector was deduced to be $\sigma_{{\rm tri}} = \sqrt{2}\sigma_{{\rm tri}-{\rm tri}}=$ 328.5(28)~ps.  Using this value, we then determined the intrinsic time resolution of the $\alpha$-ToF detector to be $\sigma_{\alpha{\rm ToF}}=\sqrt{\sigma_{{\rm tri}-\alpha{\rm ToF}}^2 - \sigma_{{\rm tri}}^2 }=$ 251.5(68) ps.

\subsection{Correlation measurement}
\par In order to make an offline determination of the $\alpha$-ToF detector performance in a correlated $E_{\alpha}$-ToF measurement, a mixed $\alpha$-source comprising  $^{244}$Cm (5804 keV), $^{241}$Am (5485 keV) and $^{237}$Np (4788 keV) was utilized. 
\par
\begin{figure}[H]
\begin{center}
\includegraphics[scale=0.43]{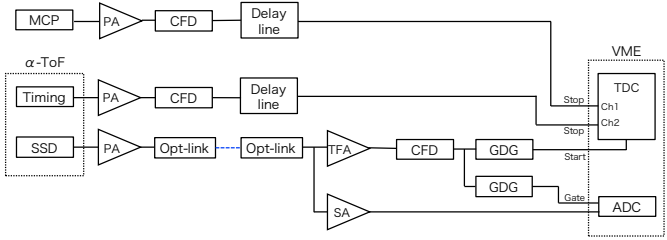}
\end{center}
\caption{Sketch of data acquisition system for the correlated $E_{\alpha}$-ToF measurement. PA: Pre Amplifier, CFD: Constant Fraction Discriminator, SA: Shaping Amplifier, TFA: Timing Filter Amplifier, GDG: Gate and Delay Generator, ADC: Analog to Digital Converter, TDC:Time to Digital Converter\label{fig6}}

\end{figure}

 
Due to its unity detection efficiency, the signal from the SSD was used to trigger the data acquisition system (Fig.\ \ref{fig6}), as the common start signal of the time-to-digital converter (TDC), and to provide $\alpha$-ray energies. Because the SSD response is slow, the timing signal of the triangle-roof detector and of the $\alpha$-ToF detector were each connected to stop inputs of the TDC through analog delay lines. The true ToF was determined from the time difference between the two stop signals.

\par
\begin{figure}[H]
\begin{center}
\includegraphics[scale=0.44]{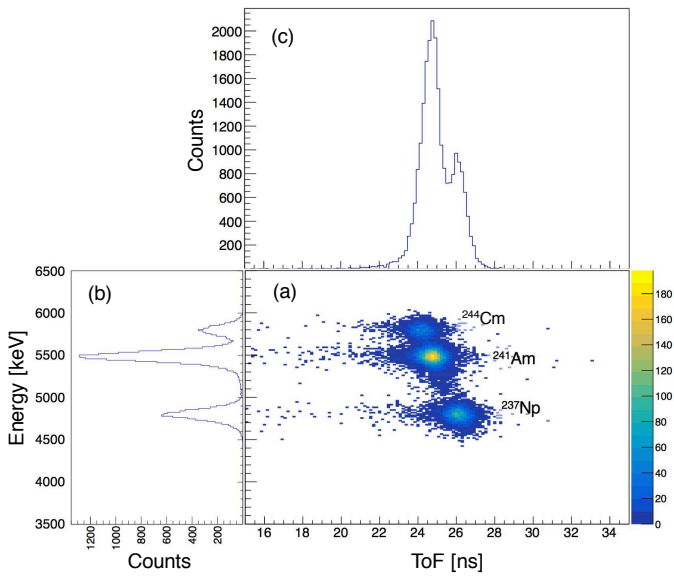}
\end{center}
\caption{(a) Two dimensional spectrum showing a correlated $E_\alpha$-ToF measurement obtained from mixed $\alpha$ source. (b)  projection onto the $E_\alpha$ axis.  (c) Projection onto the time-of-flight axis. \label{fig7} The ToF was measured from triangle-roof detector to $\alpha$-ToF detector.}

\end{figure}

 A VME based data acquisition system, with an analog to digital converter module (MADC-32) for energy signals and a TDC module (MTDC-32) for timing signals, was used. The TDC had a time resolution of 80 ps.  
Figure \ref{fig7} shows the $E_\alpha$-ToF correlated data measured using the mixed $\alpha$-source.  The projection of the two-dimensional spectrum (Fig.\ \ref{fig7} (c)) onto the horizontal axis yields the singles time-of-flight spectrum. In the singles ToF spectrum, the three different $\alpha$-ray signals are not resolved. However, they can be resolved in a correlated measurement with the energy spectrum.

\par The energy resolution for the 5.48 MeV $\alpha$-rays was determined from the energy spectrum (Fig.\ \ref{fig7} (b)) to be 141.1(9) keV, which is about twice the typical resolution of such an SSD. The degradation of the resolution is presumed to be due to noise from the MagneToF bias voltage supply.

\section{Anticipated application for nuclear spectroscopy}

\begin{figure}[H]
\begin{center}
\includegraphics[scale=0.48]{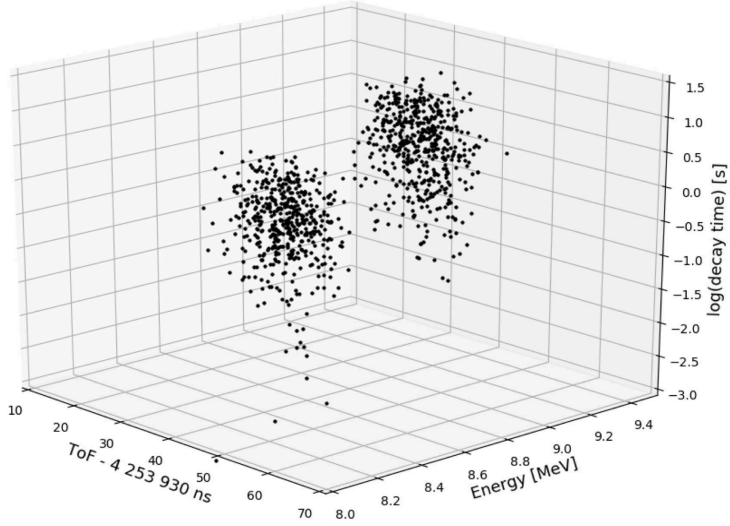}
\end{center}
\caption{Three-dimensional mapping of detection data from simulated measurement of a postulated nucleus having an isomeric state at 500~keV above the ground state and having similarly differing $\alpha$-decay energies of 8.5~MeV and 9.0~MeV for each state, with the half-lives of the ground state and isomer being 0.67 and 5.0 seconds, respectively. Information on time-of-flight, decay energies, and decay-time can be considered independently on an event by event basis. }\label{fig10}

\end{figure}

It would be desirable for the $\alpha$-ToF detector to be used for nuclear decay spectroscopy.  Table~\ref{tab1} shows candidate nuclides in the superheavy element region to be studied with a combination of an MRTOF-MS and the $\alpha$-ToF detector. 
The $\alpha$-ToF detector has the capability of correlated measurements for mass and decay properties. Using all the information -- namely time-of-flight, decay energies, and decay times -- obtained in the $\alpha$-ToF detector,  we will be able to separate the ground state from the isomeric state and perform accurate mass measurements of each. As such, we have used the above-measured performance data to perform a numerical simulation to produce a three-dimensional parameter space figure like Fig.\ \ref{fig10}.  Using the decay properties, we performed a numerical simulation that demonstrates the possibility to resolve an isomeric state in time-of-flight spectra for $^{263}$Sg and $^{267}$Hs.  

\begin{figure}[H]
\begin{center}
\includegraphics[scale=0.44]{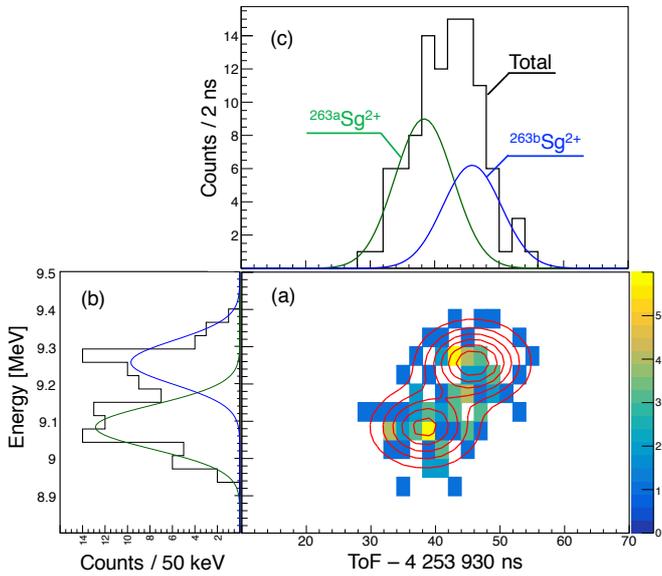}
\end{center}
\caption{Numerical simulation of $^{263}$Sg for time-of-flight and $\alpha$-energy spectra. (a) Two dimensional spectrum for the time-of-flight and $\alpha$-energy. The red lines are equal intensity contours resulting from the two-dimensional fit. 
(b) $\alpha$-energy spectrum. (c)  Time-of-flight spectrum.  Green line indicates the excited state ($^{263a}$Sg$^{2+}$) and blue line indicates the ground state ($^{263b}$Sg$^{2+}$).  \label{fig8}}


\end{figure}

Let us consider the isotope $^{263}$Sg for which two $\alpha$-decay energies, 9.06 MeV and 9.25 MeV, have been reported in the literature \cite{Folden:2004hi,Gregorich:2006gu}. However, the isomeric state and the ground state are not assigned.  Lacking contrary evidence, for our simulation we assume the excitation energy of the isomeric state to be 200 keV -- similar to the difference in reported $\alpha$-decay energies -- and the population of the two states to be equal. Figure \ref{fig8} shows the simulated results for 100 events. 
For an MRTOF-MS operating with a mass resolving power of $R_\textrm{m}$=200,000 \cite{YIto2018152501}, the ToF of $^{263a,b}$Sg$^{2+}$ peaks in the spectrum are fully unresolved (Fig.\ \ref{fig8} (c) - total). However, the $\alpha$-decay energy spectrum is sufficiently resolved so as to permit two dimensional fitting of the $\alpha$-energy and the ToF to 2 two-dimensional Gaussians as shown in Fig.\ \ref{fig8}. 
In this way, it is possible to separately produce resolved time-of-flight spectra for each state and to assign the states without ambiguity.\par

\begin{figure}[H]
\begin{center}
\includegraphics[scale=0.4]{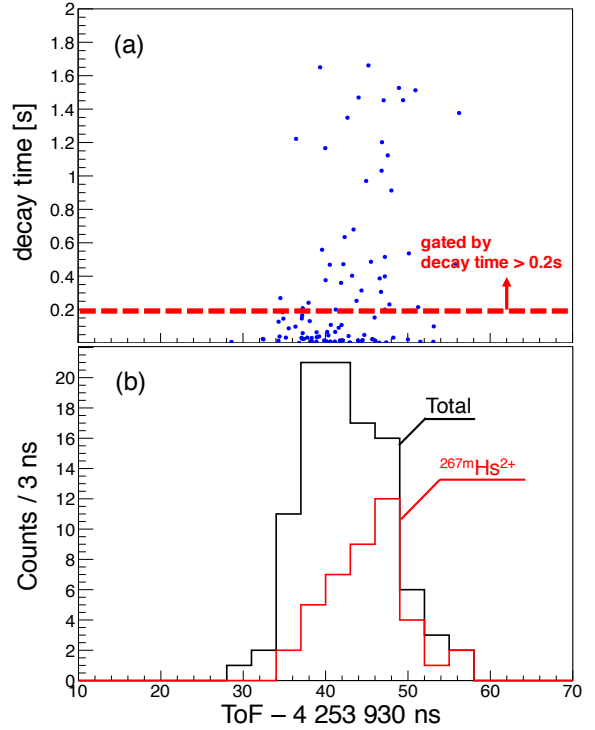}
\end{center}
\caption{Numerical simulation of $^{267}$Hs for time-of-flight and decay time. (a) Two dimensional scatter plot of time-of-flight and decay time, red dashed line indicates a lower gate line to extract isomeric state components. (b) Time-of-flight spectra without decay time gating (black) with gating on decay time $>$200  ms (red).
} \label{fig9}

\end{figure}

Let us further consider the isotope $^{267}$Hs.  In this case, the energies of the $\alpha$-rays emitted by the two states would not be resolvable by our $\alpha$-ToF detector. However, it is reported \cite{KMorita2015NRA} that the ground state and the isomeric state have greatly differing half-lives of 52$^{+13}_{-8}$~ms and 0.80$^{+3.8}_{-0.37}$~s, respectively.   Due to the large half-life difference, we could use the decay time rather than the $\alpha$-energy to distinguish the states.
Figure \ref{fig9}~(a) shows the results of a simulated measurement of $^{267}$Hs$^{2+}$ in terms of a ToF versus decay time scatter plot, where the decay time is the time interval between the ToF signal and the $\alpha$-decay signal. Within 200 ms of striking the $\alpha$-ToF detector, 94 \% of the $^{267g}$Hs will have decayed. In the same time, less than 20 \% of the $^{267m}$Hs will decay. Thus, by gating on $\alpha$-decay events occurring more than 200 ms after the ToF signal is observed it is possible to produce a ToF spectrum which is almost purified from $^{267m}$Hs ion, as shown in Fig.\ \ref{fig9}~(b).
\par


\begin{table}[H]
  \begin{center}
    \caption{Candidate nuclides for spectroscopic measurement with MRTOF + $\alpha$-ToF in the superheavy elements region. Observed $E_{\alpha}$ and the half-lives  are listed.}\label{tab1}
    \begin{tabular}{l c c c } \hline
      Nuclide & E$_{\alpha}$ & T$_{1/2}$  & Ref. \\ \hline 
      $^{267g}$Hs & 9.88 MeV & 52$^{+13}_{-8}$ ms &  \cite{ALazarev1995PRL,KMorita2015NRA} \\
      $^{267m}$Hs & 9.73 MeV & 0.80$^{+3.8}_{-0.37}$ s & \cite{ALazarev1995PRL,KMorita2015NRA} \\ \hline
      $^{262a}$Bh & 10.06 MeV & 83 $\pm$ 14 ms  & \cite{Hessberger:2009ex} \\
      $^{262b}$Bh & 10.37 MeV & 22 $\pm$ 4 ms  & \cite{Hessberger:2009ex} \\ \hline
      $^{263a}$Sg & 9.06 MeV & 0.82$^{+0.37}_{-0.19}$ s  & \cite{Folden:2004hi,Gregorich:2006gu} \\
      $^{263b}$Sg & 9.25 MeV & 310$^{+160}_{-80}$ ms  & \cite{Folden:2004hi,Gregorich:2006gu} \\ \hline
      $^{257g}$Rf & 8.78 MeV & 5.5 $\pm$ 0.4 s  & \cite{Streicher:2010ei} \\
      $^{257m}$Rf & 9.02 MeV & 4.9 $\pm$ 0.7 s &  \cite{Streicher:2010ei}\\ \hline
    \end{tabular}
    
  \end{center}
\end{table}

\section{Summary and outlook}
\par We developed an $\alpha$-ToF detector for correlated measurements of the heavy-ion time-of-flight and the decay energy.
We also developed a front-end signal processing circuit for the $\alpha$-ToF detector which is floated by a high voltage potential.
The performance test of the $\alpha$-ToF detector showed an efficiency of more than 90\% for the timing signal,  
an energy resolution of 141.1(9)~keV for 5.48~MeV $\alpha$-rays, and a time resolution of 251.5(68)~ps.
\par Although the energy resolution is somewhat degraded, it should still be sufficient to resolve many isomeric states.  Efforts to reduce the noise, in collaboration with ETP Ion Detect, are ongoing.  The measured off-line performance is sufficient to meet the initial requirements for an on-line detection system.  As such, the newly developed $\alpha$-ToF detector has been installed as the detection system in an already-commissioned MRTOF-MS for online mass measurements at the RIKENs RI-Beam Factory.  In the near future, first on-line testing will be performed using medium-yield ($\sim$0.1~s$^{-1}$) $\alpha$-emitting nuclei produced in fusion-evaporation reactions.  Following successful on-line commissioning, it is planned that first measurements with very low-yield superheavy elements will commence. 


\section*{Acknowledgments}
This work was financially supported by the Japan Society for the Promotion of Science KAKENHI
(Grant No. 17H06090). One of the author (T.N.) thanks the RIKEN Junior Research Associate Program.

\bibliography{Dev_Alpha-W10}

\end{document}